\magnification=\magstep1
\baselineskip=16pt
\def\ni{\noindent}
\def\bs{\bigskip}
\def\eq{\eqno}
\input epsf


\centerline{{\bf HADRON PHYSICS AND THE STRUCTURE OF NEUTRON
STARS}\footnote{$^{*)}$}{Lecture given at the Meeting of
Astrophysics Commission of Polish Academy of Arts and Sciences,
Cracow, June 1996.}}

\bs
\bs
\bs

\centerline{Marek Kutschera}

\bs

\centerline{H. Niewodnicza\'nski Institute of Nuclear Physics}

\centerline{ul. Radzikowskiego 152, 31-342 Krak\'ow, Poland}

\bs\bs\bs

\ni {\bf Abstract:}

The equation of state of hadronic matter in  neutron stars is    
briefly reviewed. Uncertainties  
regarding the stiffness and composition of hadronic 
matter are discussed. Importance of poorly known short range interactions of 
nucleons and hyperons is emphasized. Condensation of meson fields 
and the role of subhadronic degrees of freedom is considered. 
Empirical constraints on the equation of state emerging from observations
of neutron stars are discussed. The nature of the remnant of
SN1987A is considered.

\bs\bs\bs
 
\ni {\bf 1. Introduction}

\bs

Neutron stars are the final stage of evolution of massive stars, $M>8M_{\odot}$.
They are born in supernova explosions which terminate hydrostatic evolution 
when heavy elements up to iron are  synthesized in the
core. When the mass of still growing iron core reaches the 
Chandrasekhar limit, the  core looses stability and collapses to
a neutron star.
It is an open question what is the maximum mass of stars which leave neutron
star after the core collapse. I shall briefly mention some recent ideas  
regarding this problem.

The main subject of this lecture is the equation of state (EOS)
of hadronic matter 
which determines properties of neutron stars, in particular their internal 
structure.  Before discussing in some detail the physics
governing the EOS, I first review the observational parameters
of neutron stars.  
Then I discuss currently considered possibilities regarding the nature of the 
neutron star EOS. In the last section the empirical constraints
on the EOS are discussed. As particularly relevant,
the nature of the remnant of  SN1987A is discussed.

\bs

\ni {\bf 2. Empirical parameters of neutron stars}

\bs
Neutron stars have very small radii, on order of $10 km$, and
cool rather quickly after birth in supernova explosion. These
are principal reasons making the flux of thermal photons emitted
by neutron stars practically invisible. Only young neutron stars
which are sufficiently hot radiate X-ray flux which can be
registered by satellite-born detectors.

Isolated neutron stars which rotate fast enough and possess
strong magnetic field are observed as radio pulsars. At
present $\sim 1000$ pulsars are known and this number
continuously increases. Neutron stars are also found in
binaries.  These accreting matter from the companion are
observed as X-ray pulsars, when they have strong magnetic field,
or as X-ray bursters in case of weak magnetic field. Of
particular importance are binary pulsars with the companion
neutron star as they allow one to measure precisely the neutron star
masses. At present six such double neutron star binaries are known.

Physical parameters of neutron stars most relevant to constraining the
EOS are the mass, the radius, the surface
temperature and the age. Simultaneous determination of all these
parameters would very tightly constrain the neutron star EOS.
Unfortunately, these parameters are not easily accessible to observations.

Period of rotation and the magnetic field of pulsars could also
provide important information about the internal structure of
the stars. In particular, pulsar timing is a powerful means
of probing neutron stars.

\bs

\ni { 2.1 NEUTRON STAR MASSES}

\bs
Presently, masses of about 20 neutron stars in binary systems are
determined. Among them are six double neutron star binaries, PSR
B1913+16 [1], PSR B1534+12 [2], PSR B2303+46 [3], PSR B2127+11C
[4], PSR B1820-11 [5] and PSR J1518+4904 [6]. For three of them,
PSR B1913+16, PSR B1534+12 and PSR B2127+11C,
precise measurements of mass of both neutron stars are
available. The masses are found to be $M_1=1.44 
M_{\odot}$ and $M_2=1.39 M_{\odot}$ (PSR B1913+16), $M_1=1.34
M_{\odot}$ and $M_2=1.34M_{\odot}$ (PSR B1534+12) and $M_1=1.35
M_{\odot}$ and $M_2=1.36 M_{\odot}$ (PSR B2127+11C).
All masses lie in a rather narrow interval,
$1.3 M_{\odot}<M<1.5 M_{\odot}$. Masses of neutron stars in
X-ray pulsars are also consistent with these values although
are measured less accurately. The measured masses of neutron
stars apparently do not exceed the maximum mass which is about
$1.5M_{\odot}$.  We shall discuss possible implications of this
upper limit in the last section.

\bs

\ni {2.2 NEUTRON STAR RADII}

\bs

Radii of neutron stars are not directly observable. One can
infer, however, some plausible values from model calculations of
X-ray bursters  which are of the order of $10 km$. 

\bs

\ni {2.3 SURFACE TEMPERATURE OF NEUTRON STARS}

\bs

The X-ray flux from about 14 pulsars has been  detected [7]. The
spectrum of photons is more difficult to obtain. If measured, it
is often not consistent with the thermal emission but is rather
dominated by hard component due to magnetospheric activity. Only
for four pulsars, PSR 0833-45 (Vela), PSR 0630+18 (Geminga), PSR
0656+14 and PSR 1055-52, softer blackbody component
corresponding to surface thermal emission is determined. 

\bs

\ni {\it 2.4 THE AGE OF PULSARS}
\bs

Pulsar ages are estimated by measuring their spin down rates. 
Pulsars spin down due to conversion of rotational 
energy into radiation. A simple spin down relation is assumed,
${\dot \nu}=K\nu^n$, where $\nu$ is the rotation frequency, and $n$ is
the braking index. The constant $K$ 
for magnetic braking is proportional to $d^2/I$, where $d$ is
the magnetic moment of the star and $I$ is the moment of
inertia. If the energy loss is through radiation from a
dipolar magnetic field, the braking index is $n=3$. The
spin-down age of pulsar is then $\tau=-\nu/2{\dot \nu}$. 

The spin-down age with $n=3$ is commonly used for pulsars. Its
applicability, however, is questioned by
recent measurement of the braking index of the Vela
pulsar [8], which gives $n=1.4\pm0.2$. This value
implies that previous estimate of the age of Vela pulsar should
increase by a factor $\sim 3$. 

\bs
\ni {2.5 MAGNETIC FIELDS OF NEUTRON STARS}
\bs

Magnetic fields of radio pulsars are inferred from the spin down
relation assuming dipolar magnetic field. A striking feature is
the bimodal distribution of pulsar magnetic fields. Usual
pulsars have strong magnetic field, $B\sim 10^{12}-10^{13} G$,
whereas millisecond pulsars possess much weaker  fields,
$B\sim 10^8-10^9 G$. For some X-ray pulsars the magnetic
field is measured directly, by observation of absorption
features interpreted as cyclotron lines [9]. The values are in
the range found for normal pulsars. It should be noted that 
neutron stars in the X-ray bursters have, if any, still weaker
fields, $B<10^8 G$.

Magnetic field of neutron stars could also serve as a
probe of the neutron star EOS if its presence is determined by
the properties of dense matter.
The bimodal distribution of pulsar magnetic fields
strongly suggests existence of a magnetic phase transition in neutron
star matter [9]. There is a possibility, however, that some
component of the magnetic field of a neutron star is
inherited from the progenitor.

\bs

\ni{\bf 3. The EOS and structure of neutron stars}

\bs

Before turning to the structure of neutron stars, let us make a
more 
general remark as to the place of neutron stars among stable
astrophysical objects which include planets, normal
stars, white dwarfs and neutron stars. These four types of
stable objects differ in the nature of material whose
pressure supports them. There are, however, striking
similarities as to the physical nature of the pressure between
normal stars and white dwarfs, on one hand, and between planets
and neutron stars, on the other hand. In the former case, the
source of pressure is, respectively, the kinetic energy of
thermal plasma and degenerate electron gas, whereas in the
latter case the source of pressure is the interaction energy
of, respectively, atoms and hadrons.

The EOS
for both plasma and the electron gas is that of an ideal
gas with Boltzmann and Fermi statistics, respectively. Pressure
as a function of density and temperature is 
$p\sim \rho k_B T$, for plasma, and $p\sim \rho^{5/3}$ for
electron gas. 
Simplicity of these formulae is due
to the fact that the contribution of
Coulomb interactions is dominated by much higher kinetic
energies. The situation is opposite for condensed atomic matter
in planets and condensed hadronic matter in neutron stars. In
both systems, the interaction energy between particles
dominates, with kinetic energy playing a lesser role. This
makes calculation of the EOS a rather difficult
task. For atomic matter, the interactions are in principle known
and problems with calculating the EOS are mainly
of technical nature. For condensed hadronic matter in neutron
stars the situation is more challenging. The relevant interactions
between hadrons in dense matter are only roughly known, a fact which
is reflected in large uncertainty of the neutron star EOS. 
\bs

\ni {3.1 HADRONIC MATTER OF THE CORE OF NEUTRON STARS}
\bs

For physically relevant neutron stars with $M\sim 1.4M_{\odot}$
the core comprises most of the mass.
The neutron star core is the interior part of the neutron star,
below the crust, with densities exceeding the nuclear saturation
density, $n_0\approx 0.16 fm^{-3}$. Properties of the neutron
star crust matter will not be considered here. The crust is
composed of a crystal, which is similar to that in white dwarfs.
In the inner crust neutrons gradually fill in the space between
nuclei.  

The material, from which the neutron star core
is built of, is condensed hadronic matter, essentially in the
ground state. Just below the crust, the core matter is composed
mainly of neutrons with some admixture of protons, electrons and
muons.
This matter is condensed since the Fermi energy of
electrons is much higher than the thermal energy, $E_F>>k_BT$.
Thermal energy of particles in the neutron star is below $1
MeV$, except of first few minutes after formation.
The temperature of a newly
born neutron star is $k_BT \sim 30 MeV$. The star cools to $k_BT
\sim 1 MeV$ in a few minutes.
The electron Fermi energy in the core is $E_F \sim 100 MeV$.

Weak interactions ensure that the neutron star matter relaxes to
$\beta$-equilibrium with neutron, proton and electron chemical
potentials satisfying the condition

$$ \mu_N=\mu_P+\mu_e. \eq(1)$$

\ni When the electron chemical potential exceeds the muon rest
mass, $\mu_e \ge m_{\mu}=106 MeV$, muons appear in the matter
with the chemical potential $\mu_{\mu}=\mu_e$.

The neutron star matter is locally charge neutral, with the
lepton (electrons + muons) density equal to the proton density,

$$ n_e+n_{\mu}=n_P. \eq(2)$$

\bs

\ni {\it 3.1.1 Neutron star matter at density $n_0$}

Properties of neutron star matter of saturation density can be
inferred from empirical parameters of nuclear matter which are obtained
from nuclear mass formulae. The empirical value of nuclear
symmetry energy, $E_s=31\pm 4 MeV$ allows one to fix the proton
fraction by using Eq.(1) and (2). The proton fraction,
$x=n_P/n$, of $\beta$-stable nucleon matter of saturation density is

$$ x(n_0)\approx 0.05. \eq(3)$$

 The surface layer of the neutron star core, of density
$n_0\approx 0.16 fm^{-3}$, is the only part of the core whose
composition is determined quasi-empirically. The value
of the proton fraction, $\sim 5\%$, is known with accuracy determined
entirely by the empirical error of the nuclear symmetry energy.

\bs
\ni {\it 3.1.2 Neutron star matter beyond the saturation density}

Investigation of deeper layers of the neutron star requires
extrapolation of hadronic matter properties away from
empirically accessible domain. The lack of sufficient knowledge
of hadronic interactions at short distances makes this
extrapolation uncertain.

To obtain properties of neutron star matter at higher densities,
$n>n_0$, one has to employ a model of nucleon (hadronic)
interactions which allows one to calculate the energy density
of neutron star matter, $\epsilon$, as a function of baryon
density, $n$,  $\epsilon\equiv \epsilon(n)$. 

The pressure as a function of mass density, 
$p\equiv p(\rho)$, a relation referred to as the EOS, is
determined by the energy density, 

$$ \epsilon(n)=\epsilon_{kin}+\epsilon_{int}, \eq(4)$$

\ni where the kinetic energy density is

$$ \epsilon_{kin}={2 \over (2\pi)^3}[ \int_0^{k_N} d^3k \sqrt{k^2+m_N^2}
+ \int_0^{k_P} d^3k \sqrt{k^2+m_P^2}]. \eq(5)$$

\ni Here we use such units that $\hbar=c=1$ and the mass
density, corresponding to baryon density $n$, is
$\rho(n) \equiv \epsilon(n)$. Pressure is $p=n^2 \partial
(\epsilon/n)/\partial n$. 

Insufficient knowledge of hadronic interactions results in
uncertainty of the interaction energy density,
$\epsilon_{int}(n)$, which grows with increasing density. Various
model calculations give predictions which span quite a wide
range. Discrepancy of energy per particle for various EOS's
exceeds a factor of 2 at the same baryon density. This
translates into still higher discrepancy of pressure at a given
mass density.

Astrophysical implications of such an
uncertainty in the EOS can be best illustrated on the plot of
the density profile of a neutron star of fixed mass $M=1.4
M_{\odot}$. As one can see in Fig.1, various EOS's
give the radius of the star between $7 km$ and $15 km$. The
radius of the neutron star is thus known with a $50\%$
uncertainty. Also, as discussed below, considerable uncertainty
exists as far as the internal structure of the star is concerned.

The uncertainty of the EOS affects our knowledge
of the fundamental quantity of neutron star physics, the maximum
mass of neutron star. Various EOS's give values in
the range $1.5 M_{\odot} \le M_{max} \le 2.8 M_{\odot}$. 
The measured masses of neutron stars
require that the maximum mass for any EOS is $M_{max}>1.44
M_{\odot}$. 

We now turn to address the question of hadronic interactions and
their influence on the neutron star EOS.

\bs

\ni {3.2 HADRON INTERACTIONS AND THE EOS}
\bs

The decisive factor for deriving the EOS of
neutron star matter for densities $n \le 3n_0$ is the
interaction of nucleons. According to the theory of strong
interactions, the Quantum Chromodynamics, the nucleon-nucleon
interaction is some residual interaction between composite
objects, whose structure is determined by the primary QCD
interactions. Nucleon interactions
are not derivable at present from
the underlying theory. In this situation one must resort to
phenomenological methods. 

For calculating the neutron star EOS two main approaches are
used. In the first approach which is 
purely phenomenological,  the nucleon-nucleon interaction is 
parametrized in the form of a nonrelativistic
potential $v_{NN}$. This potential is fit to reproduce the
scattering data and 
the properties of the deuteron [10]. The second approach is based
on one-boson exchange (OBE)
model of nucleon scattering. The scattering amplitude is
calculated assuming the exchange 
of the lowest mesons, whose coupling to nucleons is adjusted to
fit the data [11]. Usually a simple Yukawa coupling of meson
fields to nucleons is used.
 
EOS's obtained in both approaches differ considerably. In fact
one obtains two distinct classes of EOS depending on the
way in which the nucleon-nucleon interaction is constructed. 

The energy per particle as a function of density,

$$ E(n)={1 \over n} (\epsilon_{kin}+\epsilon_{int}), \eq(6)$$

\ni in either case 
is obtained by solving many-body theory with a given model
interaction. 

Any realistic EOS should reproduce
empirical parameters of nuclear matter, which include the
saturation density, $n_0=0.16\pm 0.015 fm^{-3}$, the binding energy,
$E(n_0)-m=-16\pm 0.2 MeV$, the compressibility modulus,
$K_V=220 \pm 30 MeV$ and the symmetry energy, $E_s=31\pm 4MeV$.

Despite the fact that realistic nuclear interactions reproduce the above
saturation properties, they give different predictions at higher
densities. Of particular importance for neutron star physics are
such high density properties of the EOS as
the stiffness and the proton content. One can identify the 
components of the nuclear potential responsible for these
quantities, which are, respectively, the central potential,
$v_c(r)$, and the isospin potential, $v_{\tau}(r)$. Retaining only
these components, the nucleon-nucleon potential is

$$ v_{NN}(r) = v_c(r) + v_{\tau} {\vec \tau}_1{\vec \tau}_2+... \eq(7)$$

The short-range behaviour of the central potential, $v_c(r)$,
governs mainly the stiffness of the equation of state at high
densities. At short distances, the nucleon potential possesses a
repulsive core. The harder the core, the stiffer is the EOS.

The isospin potential, $v_{\tau}$, determines the proton fraction
of neutron star matter at high densities. For many phenomenological
potentials, such as Reid's potential [12], Urbana $v_{14}$ [10], Argonne
$v_{14}$ [13], the isospin potential is negative, $v_{\tau}<0$,
and decreases at short distances. 
In this case, the potential (7) between proton and neutron
is more repulsive than between two neutrons. Consequently, it is
energetically favourable for protons to disappear at high
densities [14,15,16]. In Fig.2 we show the isospin potential
$v_{\tau}(r)$ corresponding to some of the above interactions.

Models based on the OBE potentials, on the other
hand, predict positive isospin potential, $v_{\tau}>0$, which
increases at short 
distances. In this case, the potential (7) between proton and neutron is
less repulsive than that for a pair of neutrons. We thus expect
an increase of the proton fraction of neutron star matter at
high density. 

In Fig.3 we show the proton fraction of neutron star matter for
both classes of models.
A general tendency is that models of dense matter based on
phenomenological potentials, which have $v_{\tau}<0$, predict
the proton fraction of the order of a few percent which
decreases at high densities [14,15,16]. Models employing
the OBE interactions predict the opposite behaviour, the proton
fraction monotonically increases with density [16,17]. It is obvious
that one class of models is wrong, however, at present we are not
able to discriminate between them. The  empirical way to do so
would be to device an experiment sensitive to the short range
proton-neutron interaction. 

\bs

\ni {3.3 NEW DEGREES OF FREEDOM}
\bs

At densities above $\sim 3n_0$ one must consider 
possibility that new degrees of freedom appear in the system.
These include heavier baryons (hyperons) and condensates of
meson fields. 
At still higher densities a phase transition to quark matter
could also occur.

\bs

\ni {\it 3.3.1 Hyperons}

Possibility that hyperons become abundant in the neutron star
matter stems from the fact that at high enough density the
neutron Fermi energy can exceed the rest mass
of  hyperon $\Lambda$, which is $m_{\Lambda}=1116 MeV$ and at
still higher density
the mass of hyperon $\Sigma$, $m_{\Sigma}=1190 MeV$. 

The Fermi sea of a given neutral hyperon species starts to be populated
at the threshold density, $n_{th}$, when the neutron chemical
potential for the first time becomes equal to the hyperon chemical 
potential in the neutron star matter,

$$ \mu_N(n_{th}) = \mu_H(n_{th},n_H=0). \eq(8)$$

\ni The threshold condition for negatively charged hyperons is

$$ \mu_N(n_{th})+\mu_e(n_{th}) = \mu_H(n_{th},n_H=0). \eq(9)$$

\ni In these formulae the hyperon density at the
threshold is $n_H=0$.

Conditions (8) and (9)
are  very sensitive to interactions of the hyperon with
neutrons. The hyperon chemical potential at the threshold
can be written as

$$ \mu_H \approx m_H+ E_{int}^{NH}(n_{th}), \eq(10) $$

\ni where the last term represents the interaction energy of a
single hyperon with the neutron star matter. 

Sometimes a misleading argument is used to show that
hyperon component is present at high density.
A simple estimate of the threshold density can be obtained if one
ignores the interaction term in Eq.(10). With this assumption
the condition (8) can 
always be satisfied at some  density, as the neutron
chemical potential monotonically increases with density.
However, such an estimate can be meaningless, as the
the interaction term cannot be neglected.

Our knowledge of hyperon interactions in high density matter is
insufficient to conclude that hyperon Fermi sea is present in the
neutron star matter. 
The strength of the hyperon-neutron interaction relative to the
neutron-neutron interaction will decide whether hyperons appear
in the neutron star matter or not. In terms of the short-range potentials,
if the neutron-hyperon interaction is more repulsive than the
neutron-neutron one,

$$ v_{NH}>v_{NN}, \eq(11)$$

\ni no hyperons appear in dense matter at any density [14]. If the
opposite relation is true at short distances,

$$ v_{NH} \le v_{NN}, \eq(12)$$

 \ni there can exist a threshold density, above which the
hyperons appear in the matter [18].  

The situation here is analogous to that for protons discussed
above where also the relative strength of proton-neutron and
neutron-neutron interaction determines the proton content at
high densities.

\bs

\ni {\it 3.3.2 Meson condensates}

Attractive interactions of light mesons with nucleons in the
neutron star matter can lower
their effective mass sufficiently enough that formation of
the Bose condensate could be energetically favourable.
Negatively charged mesons, $\pi^-$ and $K^-$, can be formed when
their chemical potentials become equal to the electron chemical
potential, 

$$ \mu_{\pi^-}=\mu_e,~~~~~~~~~~~~~~~ \mu_{K^-}=\mu_e. \eq(13) $$

The meson chemical potential is the lowest eigenstate of the
meson in the matter. Once the threshold condition (13) is
satisfied, the lowest mode becomes macroscopically populated. 

The possibility of pion condensation was extensively discussed
in the literature [19]. The p-wave pion-nucleon interaction results
in a spatial modulation of the condensate. Presence of the pion
condensate strongly affects cooling of neutron stars. Because of
additional binding, the EOS of pion condensed matter is softer. 

Recently possibility of the kaon condensation in neutron star matter
has been considered [20]. The kaon-nucleon interaction is mainly
s-wave
and the condensate is uniform [21]. Neutron star matter with kaon
condensation has rather soft equation of state. A very
interesting feature of this EOS is the fact that maximum neutron
star mass calculated for cold matter, after the neutrinos
trapped in the newly born neutron star are radiated away,
is lower than the one with trapped neutrinos [21].

\bs

\ni {\it 3.3.3 Quark matter}

It is expected that quarks which are constituents of hadrons 
become liberated at high densities. The simple argument
is of geometrical nature: Hadrons are extended objects which at
high densities must overlap deconfining the quarks. The
deconfinement density is strongly model dependent. Estimates
give the deconfinement phase transition in the range $4 - 10
n_0$. Maximum quark core corresponds to the continuous phase
transition from nucleon matter to quark matter [22].

This picture of hadron to quark phase transition may be
somewhat naive in view of recent lattice QCD calculations.
At high densities and/or temperatures QCD predicts restoration
of chiral symmetry.  It is
important for our picture of dense matter in neutron star
which transition occurs first. Common assumption is that the
deconfinement occurs at lower densities than the restoration of
chiral symmetry [23,24]. Quark matter relevant to neutron stars
with broken chiral symmetry can develop spatially modulated
chiral condensate [23,24]. 

Recently a different scenario of hadron to quark transition has been
considered by G. Brown et al.[25] who propose that at high
temperatures and/or densities hadrons become first massless and
the color deconfinement occurs at much higher temperature and/or
density.

\bs

\ni {3.4 UNCERTAINTY OF EOS AND THE STRUCTURE OF NEUTRON STARS}
\bs

Theoretical models of dense matter discussed above 
constrain  weakly  stiffness of the EOS which is the most
important property of the neutron star EOS. 

\bs

\ni {\it 3.4.1 Stiffness of EOS and the maximum mass of neutron star}

Stiffness is a global property of the EOS which measures how
fast the pressure increases with increasing mass density. In
Sect.3.2 we discussed the influence of the repulsive core in
nucleon-nucleon potential on the stiffness. Also presence of
hyperons and meson condensates in the neutron star matter
affects stiffness of EOS, which becomes softer. A similar
softening of the EOS occurs when the quark phase is present. 

The stiffness of the neutron star EOS determines the value of
the maximum mass of 
neutron star, which plays the role of Chandrasekhar mass for
neutron stars. 
A general rule is that the maximum mass for soft EOS
is lower than that for stiffer EOS,
$M_{max}^{soft}<M_{max}^{stiff}$. Softest EOS's, still compatible with
measurements of neutron star masses, give $M_{max}\approx 1.5
M_{\odot}$. Stiffest realistic EOS's predict $M_{max}\approx 2.8
M_{\odot}$. 

\bs

\ni {\it 3.4.2 Structure of neutron stars}

Internal structure of the neutron star depends on its mass. Here
we consider physically relevant neutron star of mass $M=1.4
M_{\odot}$. 
 For a neutron star of a given
mass the radius for soft EOS is smaller than for stiff EOS,
$R^{soft}<R^{stiff}$. The central density of
this star is higher in case of soft EOS than for stiff EOS,
$n_c^{soft}>n_c^{stiff}$. Also, the crust thickness for soft EOS
is lower than for stiff EOS.

The proton fraction of neutron star matter has important
consequences for magnetic properties. If proton fraction is low,
$x\sim 0.05$,
and does not increase with density, it is likely that protons
become localized [26,27] at high densities. Localized protons
can form a crystal lattice [28]. Neutron star matter with
localized protons 
is unstable with respect to spontaneous polarization [29]. This
phase of neutron star matter possesses permanent magnetization
[29,30] and can contribute to the magnetic moment of the neutron star.
The same proton-neutron interactions which are responsible for
low proton fraction of uniform matter, tend to separate protons
and neutrons [31] and localize the protons.

When proton fraction increases with density, it can exceed the
threshold for direct URCA process, $x_{URCA}\approx 0.11$. This would
strongly enhance the cooling of neutron stars.

Available EOS's allow us to construct various scenarios of the
internal structure of the neutron star core. Let us briefly
describe two consistent possibilities, corresponding to two
classes of neutron star EOS's, considered in Sect.3.2. 

The models of EOS based on phenomenological potentials, which
predict low proton fraction, 
suggest the following structure. Below the solid crust there is a
layer of normal 
uniform matter, with proton fraction $\sim 5\%$. At some deeper
level, where the 
localization density is exceeded, proton localization occurs and
neutron star matter acquires 
magnetization [30]. There exist thus an inner shell which is
magnetized. If the transition to quark matter occurs at neutron
star densities, there could exist a quark core surrounding the
center of the star. 

In this scenario it is unlikely that hyperons exist in the
neutron star matter as they are expected to interact with
neutrons in a similar way as do the protons. Also, the chemical
potential of electrons decreases with density making kaon
condensation rather unlikely. 

The class of EOS's based on OBE potentials or relativistic
mean field theory predicts quite a different internal structure
of the neutron star. The layer, just below the crust, is the
same as for phenomenological models. It is
normal uniform matter containing some $5\%$ of protons. However,
in deeper layers the proton fraction increases and hyperons
appear in the neutron star matter. In still deeper layers kaon
condensate is present. It is likely in this scenario that the
chiral symmetry is restored in the center of the star. Neutron
star matter near the center is composed of a mixture of nearly
massless hadrons. 

There exist a variety of neutron star EOS's
derived by astrophysicists. Clearly, some of them are
incompatible with one another. Below we discuss how the neutron
star EOS could be constrained empirically. 

\bs

\ni {\bf 4. Empirical constraints on the neutron star EOS}
\bs

The EOS of neutron star matter at high densities is not at
present constrained sufficiently by theory to allow conclusive
statements as to the internal  structure of neutron stars. In
view of weakness of theoretical constraints it is urgent to
empirically constrain the EOS.

\bs

\ni {\it 4.1 LABORATORY EXPERIMENTS}
\bs

Scattering of heavy ions, which is the only laboratory way to
study properties of dense matter, does not probe directly the
EOS relevant to neutron stars. In nuclear collisions highly
excited hadronic matter is formed which decays quickly into
stable particles. One should perform extrapolations in order
to obtain ground state properties from  these data which would
involve considerable uncertainty. However, many important
informations regarding the neutron star EOS can be inferred from
scattering data. In particular, interactions of hyperons formed
in nuclear collisions with nucleons in dense fireball can be
studied. Also, detection of quark-gluon plasma in heavy ion
collisions could give valuable information about
energy density range in which one can expect deconfinement
transition in neutron star matter.

\bs

\ni {\it 4.2 OBSERVATIONS OF NEUTRON STARS}
\bs

Discovery of a sub-millisecond pulsar would severely constrain
the EOS. Unfortunately, reported observations of 0.5 ms 
pulsar in the SN1987A remnant [32] turned out to be erroneous.
The fastest millisecond pulsar of period 1.56 ms does not
exclude any realistic EOS.

\bs 

\ni {\it 4.2.1 X-rays from rotation-powered pulsars}

Observation of thermal flux of photons from a pulsar, whose age
can be estimated, can provide information how fast  the
neutron star cools. Probing the cooling curve of neutron
stars is considered to be the most profitable method to learn
about the internal composition of neutron stars.  

Recent observations of thermal X-ray flux from four pulsars give
promise that in near future the cooling curve will be
empirically constrained. The main objective of these
observations is to discriminate between fast and slow cooling
mechanisms. 

Slow cooling proceeds mainly through the modified
URCA process,

$$ n+n \to n+p+e^-+{\bar \nu_e},~~~~~~~ n+p+e^- \to
n+n+\nu_e. $$

\ni It is the dominating  mechanism of standard cooling
scenario for neutron stars
whose proton content is below the critical value, $x<x_{URCA}
\approx 0.11$.

If the proton content exceeds the critical value, $x>x_{URCA}$, or
there exist kaon (or pion) condensate in neutron star matter,
cooling proceeds through direct URCA process,

$$ n \to p+e^-+{\bar \nu_e},~~~~~~~ p+e^- \to n+\nu_e. $$

\ni This cooling mechanism is much faster, and, correspondingly,
the temperature of the neutron star is lower than for modified
URCA. 

Recent comparison of the X-ray luminosities of four pulsars [33]
is not conclusive, but the observational data are somewhat
closer to the standard cooling curve.

\bs

\ni {\it 4.2.2 Remnant of SN1987A}

Detection of neutrino flux associated with optical observation
of supernova SN1987A was the best confirmation of the theory of
neutron star formation in supernova Type II explosions. Present
observations of the light curve of the remnant of SN1987A do not
confirm existence of the neutron star. Continuously decreasing
luminosity of the remnant of SN1987A suggests that Crab-like
pulsar does not exist in the remnant. Also, no hot X-ray source
is observed. This lack of signature of neutron star led some
authors [34] to speculate that the neutron star formed initially
in SN1987A was in a metastable state and subsequently collapsed
to black hole. Presence of a black hole in SN1987A would strongly
constrain the neutron star EOS.
 
Scenario of black hole formation in SN1987A is as follows. The
progenitor star of SN1987A is known to have mass
$18M_{\odot}<M<20M_{\odot}$. Evolutionary calculations show [35]
that this star developed an iron core of mass $\sim 1.6 M_{\odot}$
which collapsed to form hot neutron star. This neutron star
existed at least for $\sim 10s$, a period when the neutrino
emission took place. After radiating away the trapped neutrinos
the neutron star has lost stability and collapsed to black hole.

This scenario requires that the EOS has some unique features.
The maximum mass corresponding to the hot neutron star matter,
with trapped neutrinos, $M_{max}^{hot}$, has to be higher than the
mass $M_{ns}\sim 1.6 M_{\odot}$ of the
neutron star formed in the collapse of SN1987A,
$M_{ns}<M_{max}^{hot}$.  After emitting neutrinos
the neutron star looses stability which requires that
its mass is higher than the maximum mass of a neutron star
corresponding to cold neutron star matter, $M_{max}^{cold}<M_{ns}$.
The EOS with kaon condensation can meet these constraints [21],
as we mentioned in Sect.3.3.2. The maximum mass for cold EOS is
$M_{max}^{cold}=1.5 M_{\odot}$.

The maximum mass of neutron star of $1.5 M_{\odot}$ explains in
a natural way the cutoff observed in measured masses of neutron
stars. Existence of such a limit is very surprising in view of
the fact that considerable amount of material, at least a few
tenths of solar mass, is expected to fall back onto newly formed
neutron star after the explosion. One would expect many heavier
neutron stars to be formed.

If the maximum mass of neutron star is $M_{max}=1.5 M_{\odot}$
then one can determine maximum mass of the progenitor star,
whose collapse can leave the neutron star remnant. For an
isolated star this mass is about $20 M_{\odot}$. Heavier stars
are expected to leave black hole remnants. The problem of formation of
many low mass black holes in Galaxy is discussed in [36].

\bs
This work is partially supported by the Polish State Committee
for Scientific Research (KBN), grants 2 P03D 001 09 and 2 P03B
083 08.

\bs\bs

\ni {\bf References}

\bs
\ni~1.~~ J. H. Taylor and J. M. Weisberg, Astrophys. J. {\bf
345}, 434 (1989).

\ni~2.~~A. Wolszczan, Nature {\bf 350}, 688 (1991).

\ni~3.~~S. E. Thorsett, Z. Arzoumanian, M. M. McKinnon and J. H.
Taylor, Astrophys. J. 

\ni~~~~~{\bf 405}, L29 (1993).

\ni~4.~~W. T. S. Deich and S. R. Kulkarni, in {\it Compact Stars
in Binaries}, J. van Paradijs, 

\ni~~~~~E. P. J. van den Heuvel and E.
Kuulkers eds., Dordrecht, Kluwer (1996).

\ni~5.~~Z. Arzoumanian, Ph. D. thesis, Princeton Univ. (1995).

\ni~6.~~D. J. Nice, R. W. Sayer and J. H. Taylor, Astrophys. J.
{\bf 466}, L87 (1996).

\ni~7.~~H. \"Ogelman, in {\it The Lives of Neutron Stars}, M. A.
Alpar, U. Kiziloglu and J. van 

\ni~~~~~Paradijs eds., Dordrecht, Kluwer
(1995). 

\ni~8.~~A. G. Lyne, R. S. Pritchard, F. Graham-Smith and F.
Camilo, Nature {\bf 381}, 497 (1996).

\ni~9.~~K. Makishima, in {\it The Structure and Evolution of
Neutron Stars}, D. Pines, R. Tamagaki 

\ni~~~~~and S. Tsuruta eds.,
Addison-Wesley (1992).

\ni 10.~~I. E. Lagaris and V. R. Pandharipande, Nucl. Phys. {\bf
A359}, 331, 349 (1981).

\ni 11.~~R. Brockmann and R. Machleidt, Phys. Rev. C  {\bf 42},
1965 (1990).

\ni 12.~~R. V. Reid, Ann. Phys. {\bf 50}, 411 (1968).

\ni 13.~~R. W. Wiringa, R. A. Smith and T. L. Ainsworth, Phys. Rev. C
{\bf 29}, 1207 (1984).

\ni 14.~~V. R. Pandharipande and V. K. Garde, Phys. Lett. {\bf
39B}, 608 (1972).

\ni 15.~~R. W. Wiringa, V. Fiks and A. Fabrocini, Phys. Rev. C
{\bf 38}, 1010 (1988).

\ni 16.~~M. Kutschera, Z. Phys. A {\bf 348}, 263 (1994).

\ni 17.~~K. Sumiyoshi, H. Toki and R. Brockmann, Phys. Lett.
{\bf B276}, 393 (1992).

\ni 18.~~N. K. Glendenning, Nucl. Phys. {\bf A493}, 521 (1989).

\ni 19.~~G. E. Brown and W. Weise, Phys. Rep. {\bf 27C}, 1 (1976).

\ni 20.~~D. B. Kaplan and A. E. Nelson, Phys. Lett. {\bf B175},
57; {\bf B179}, 409(E) (1986).

\ni 21.~~V. Thorsson, M. Prakash and J. M. Lattimer, Nucl. Phys.
{\bf A572}, 693 (1994).

\ni 22.~~M. Kutschera and A. Kotlorz, Astrophys. J. {\bf 419},
752 (1993).

\ni 23.~~M. Kutschera, W. Broniowski and A. Kotlorz, Phys. Lett.
{\bf B237}, 159 (1990).

\ni 24.~~M. Kutschera, W. Broniowski and A. Kotlorz, Nucl. Phys.
{\bf A516}, 566 (1990).

\ni 25.~~G. E. Brown, A. D. Jackson, H. A. Bethe and P. M.
Pizzochero, Nucl. Phys. {\bf A560}, 

\ni~~~~~1035 (1993).

\ni 26.~~M. Kutschera and W. W\'ojcik, Acta Phys. Pol. {\bf
B21}, 823 (1990).

\ni 27.~~M. Kutschera and W. W\'ojcik, Phys. Rev. C {\bf 47},
1077 (1993).

\ni 28.~~M. Kutschera and W. W\'ojcik, Nucl. Phys. {\bf A581},
706 (1994).

\ni 29.~~M. Kutschera and W. W\'ojcik, Phys. Lett. {\bf B223}, 1
(1989).

\ni 30.~~M. Kutschera and W. W\'ojcik, Acta Phys. Pol. {\bf
B23}, 947 (1992).

\ni 31.~~M. Kutschera, Phys. Lett. {\bf B340}, 1 (1994).

\ni 32.~~C. Kristian, et al., Nature {\bf 338}, 234 (1989).

\ni 33.~~Ch. Schaab, F. Weber, M. K. Weigel and N. K.
Glendenning, astro-ph/9603142.

\ni 34.~~G. E. Brown, S. W. Bruenn and J. C. Wheeler, Comments
Astrophys. {\bf 16}, 153 (1992).

\ni 35.~~S. E. Woosley and F. X. Timmes, astro-ph/9605121.

\ni 36.~~G. E. Brown, J. C. Weingarten and R. A. M. J. Wijers,
astro-ph/9505092.

\bs


\bigskip
\bigskip

\epsffile{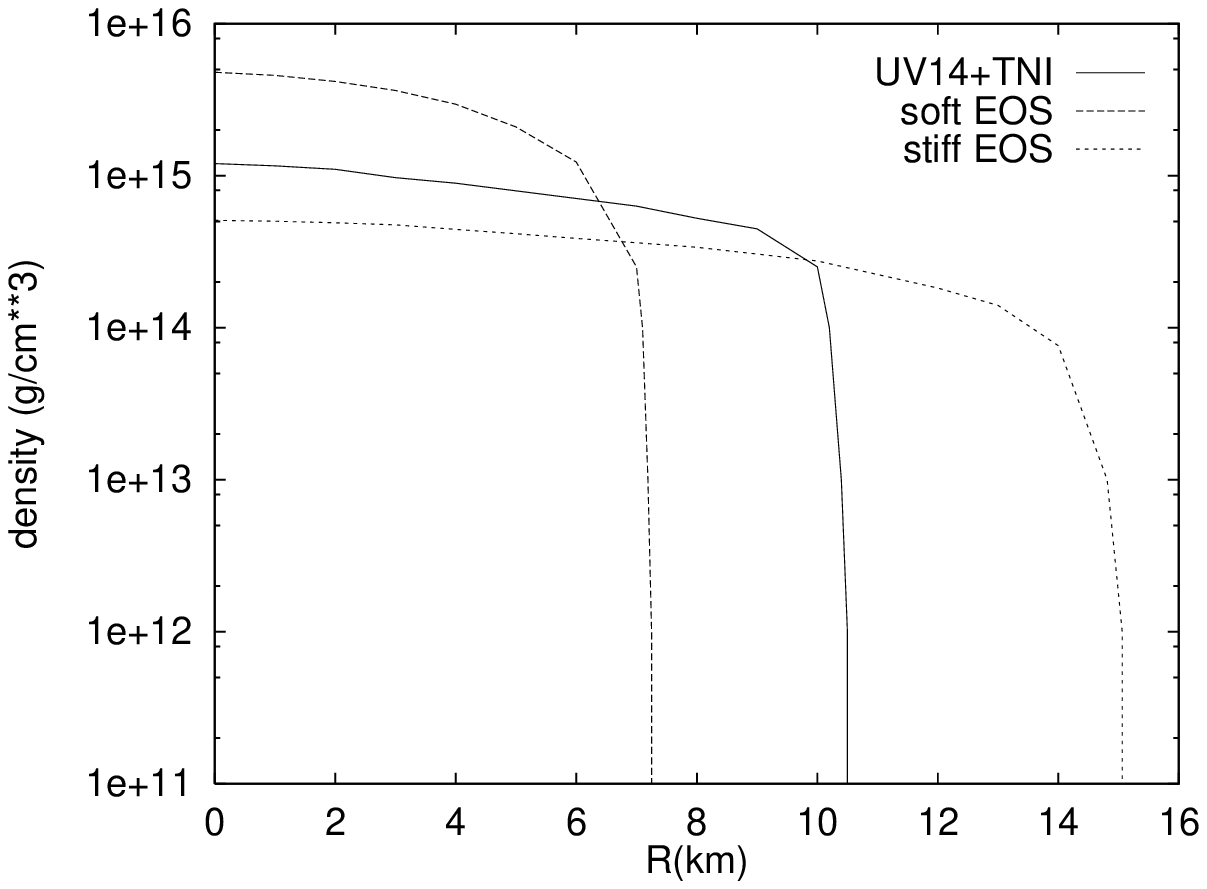}

\noindent Fig.1

\noindent Density profile of $1.4 M_{\odot}$ neutron star for typical
soft, medium, and stiff equations of state.

\bigskip
\bigskip
\vfill
\break

\epsffile{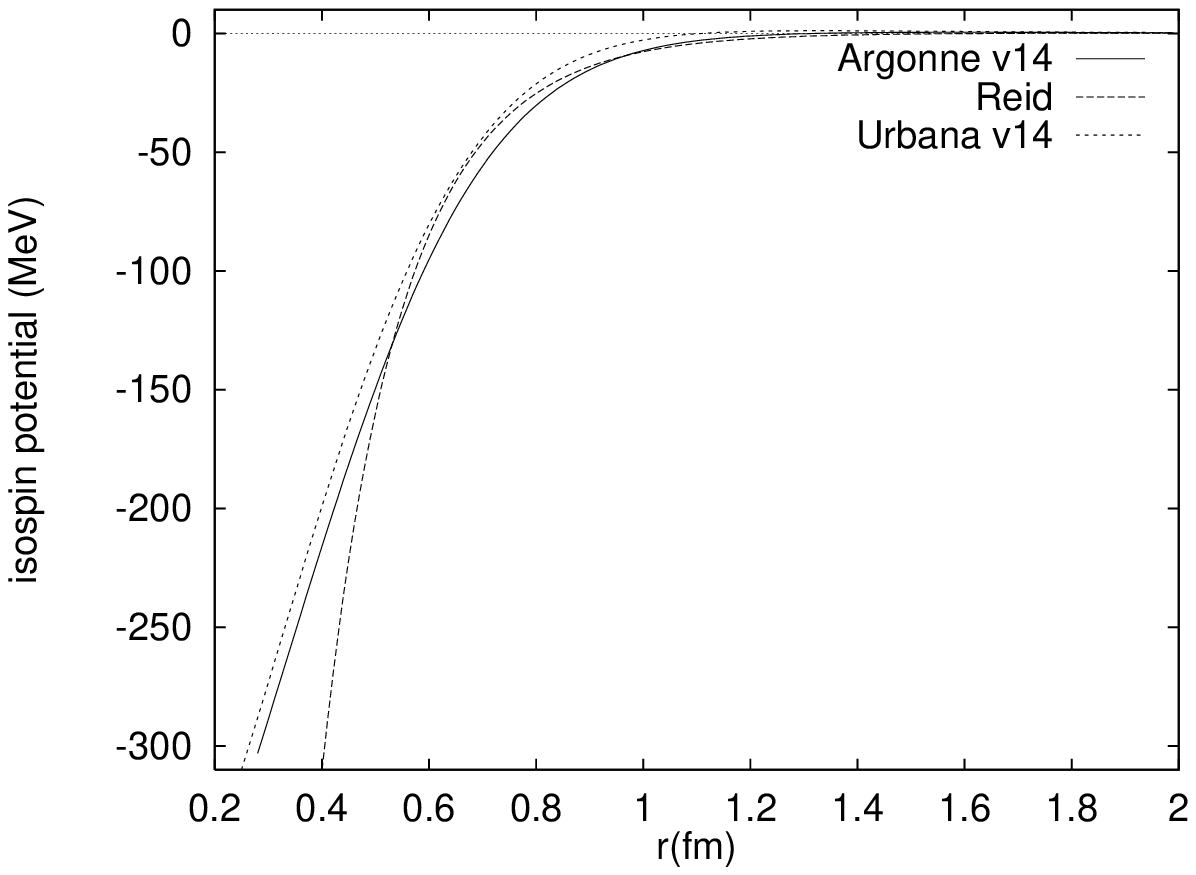}
\noindent Fig.2

\noindent The isospin potential of  Reid, Argonne $v_{14}$, and Urbana
$v_{14}$ nuclear potentials.

\bigskip
\bigskip

\vfill
\break
\epsffile {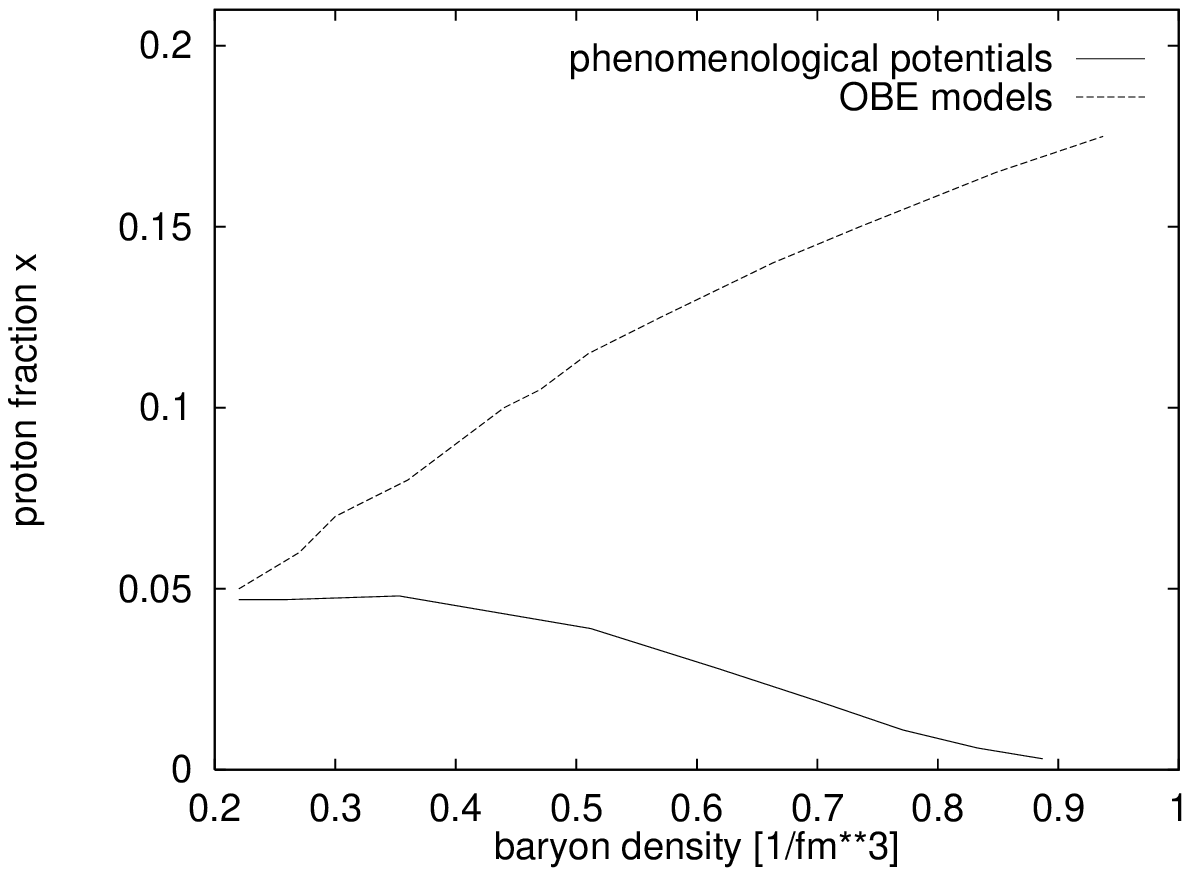}
\noindent Fig.3

\noindent Proton fraction as a function of density for OBE models and for
phenomenological models with isospin potential $v_{\tau}<0$.

\end